\documentclass[aps,prb,10pt,twocolumn,showpacs,floatfix,a4paper]{revtex4}
\usepackage{graphicx}
\usepackage[intlimits]{amsmath}

\begin{document}

\title{Capacitance of a resonant level coupled to Luttinger liquids}
\author{Moshe Goldstein}
\author{Richard Berkovits}
\affiliation{The Minerva Center, Department of Physics, Bar-Ilan
University, Ramat-Gan 52900, Israel}

\begin{abstract}
In this paper we study the differential capacitance of a single-level
quantum dot attached to quantum wires in the Luttinger liquid phase,
or to fractional quantum Hall edges,
by both tunneling and interactions.
We show that logarithmic or even power-law
divergence of the capacitance at low temperatures
may result in a substantial region of the parameter space
(including the $\nu=1/3$ filling).
This behavior is a manifestation of generalized two channel Kondo physics,
and can be used to extract Luttinger
liquid parameters from thermodynamic measurements.
\end{abstract}


\maketitle

\section{introduction} \label{sec:intro}
The study of low-dimensional strongly-correlated systems
has been at the focus of numerous experimental and theoretical
efforts in the last decades.
Two important categories in this field are 
quantum impurities,
systems of a few interacting degrees of freedom coupled
to noninteracting environments (e.g., 
the spin-boson, 
Anderson, and Kondo models \cite{hewson});
as well as gapless one-dimensional systems, whose 
low energy physics
is governed by the Luttinger liquid (LL) theory
\cite{bosonization}.
With recent advance in fabrication techniques,
quantum impurities and LLs are becoming the basic building-blocks
of nanoelectronic circuits:
both can be realized in semiconductor heterostructures,
metallic nanograins and nanowires, or carbon-based materials.
LL physics also applies
to fractional quantum Hall effect (FQHE) 
edges \cite{chang03}.
Research of circuits composed of components
from \emph{both} families thus not only allows us to
study quantum impurities coupled to nontrivial environments
in order to shed light on both systems,
but also is of much experimental relevance.
There is thus no wonder that such problems have
attracted significant attention recently.
However, most of it was devoted to transport properties
\cite{bosonization,chang03,kane92,
nazarov03,komnik03,goldstein10b}, 
while other (e.g., thermodynamic) phenomena received much less treatment
\cite{furusaki02,kolomeisky02,kakashvili03,lehur05,sade05,wachter07,
bishara08,goldstein09}.
This situation is now rapidly changing, with a surge
of interest in the capacitance of low-dimensional systems
\cite{buttiker93,berman99,gabelli06,hamamoto10,mora09}.

\begin{figure}[b]
\includegraphics[width=4.5cm,height=!]{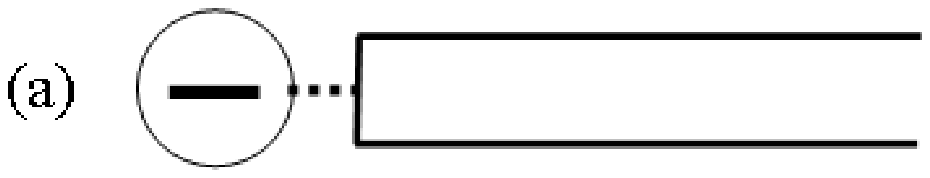}
\vskip 0.1cm
\includegraphics[width=7cm,height=!]{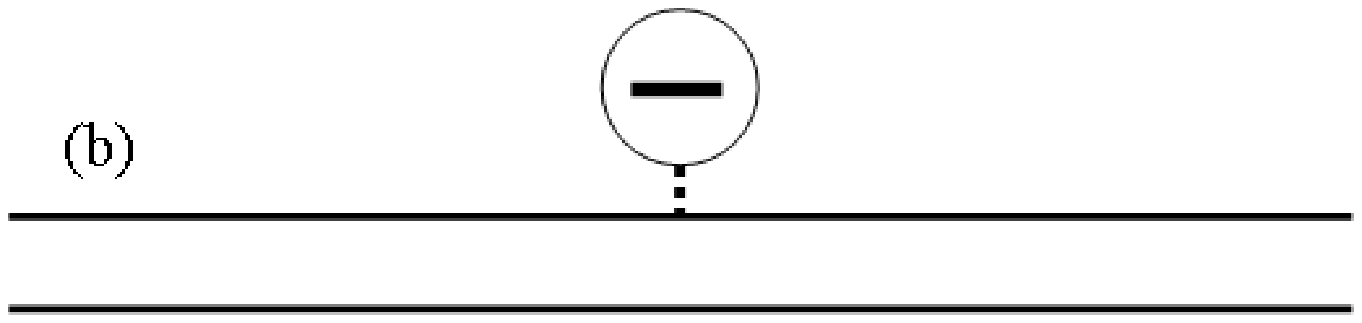}
\vskip 0.15cm
\includegraphics[width=7cm,height=!]{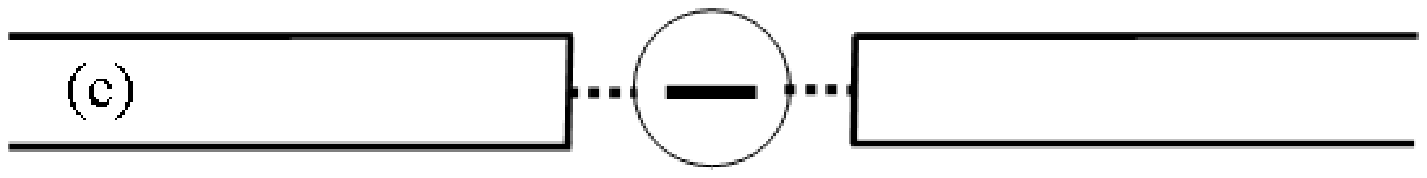}
\vskip 0.15cm
\includegraphics[width=7cm,height=!]{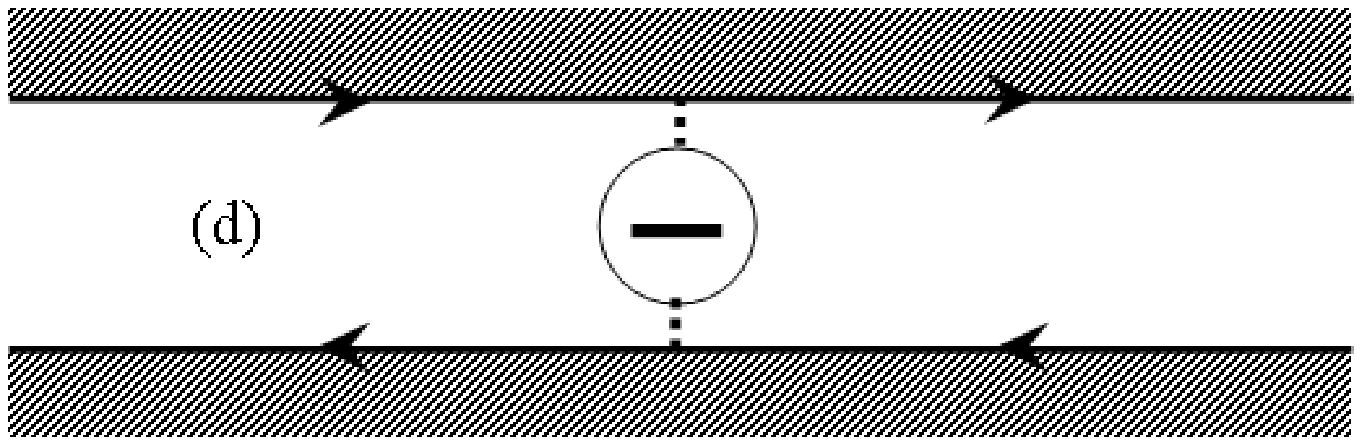}
\caption{\label{fig:geometry}
Different configurations of a level coupled to LL wires:
(a) end-coupled;
(b) side-coupled;
(c) embedded;
(d) level coupled to FQHE edges (with FQHE bars hatched).
}
\end{figure}

Previous studies \cite{furusaki02,lehur05,goldstein09}
have addressed the static differential capacitance of
a level coupled to a single FQHE edge,
or, equivalently \cite{gogolin94},
to the end of a single quantum wire,
cf.\ Fig.~\ref{fig:geometry}(a).
This problem was found to be equivalent to an anisotropic single-channel
Kondo model, where the occupied and unoccupied 
states of the level correspond to the two spin
states of the Kondo impurity.
The system can thus be in one of two phases:
(i) An antiferromagnetic-Kondo type phase,
in which the level is strongly-coupled
to its environment at low energies,
leading to a constant low temperature static capacitance;
(ii) A ferromagnetic-Kondo type phase, where the level
is decoupled at low energies, leading to a $\sim 1/T$ divergence
of the capacitance when the level energy crosses the the Fermi energy.
Moreover, the dependence of the capacitance on the various interactions
in the system is universal
\cite{goldstein09}, so it cannot be used to extract LL behavior.

In this work we thus proceed to investigate
intrinsically multi-channel scenarios,
such as the side-coupled and embedded geometries, depicted in
Figs.~\ref{fig:geometry}(b) and \ref{fig:geometry}(c), respectively.
An equivalent system is a quantum dot tunnel-coupled to two
FQHE edges [Fig.~\ref{fig:geometry}(d)].
A gate electrode can be used both to control the level
energy and to measure the capacitance through AC conductance \cite{gabelli06};
the capacitance could also
be probed using a quantum point contact charge sensor. 
We find that in these configurations there are regimes of power-law
behavior of the differential capacitance at low energies.
Hence, we have here the rare opportunity of experimentally
demonstrating LL behavior and extracting LL parameters
from thermodynamic measurements.
Furthermore, we show how these phenomena
expose generalized two-channel Kondo (2CK) physics \cite{hewson},
and may thus enable its experimental realization.
We introduce our model in Sec.~\ref{sec:model},
present a preliminary analysis in Sec.~\ref{sec:prelim}
and a general treatment in Sec.~\ref{sec:analysis},
and summarize our findings in Sec.~\ref{sec:conclude}.

\section{Model} \label{sec:model}
We consider a single level quantum dot embedded between the ends
of two LL wires [cf.\ Fig.~\ref{fig:geometry}(c)]
or between two FQHE edges [Fig.~\ref{fig:geometry}(d)],
and use the terminology ``leads''
(denoted by $\ell=L,R$) to refer to both possibilities.
Assuming spin-polarized electrons
(the natural situation in the FQHE case),
both systems are then described by the Hamiltonian
$H=H_D+H_{W}+H_T+H_U$. 
$H_D=\varepsilon_0 d^{\dagger} d$ is the level Hamiltonian,
with $d$ the level Fermi operator, and $\varepsilon_0$ its energy.
The bosonized Hamiltonian of the leads is
$H_W = v/(4\pi) 
\sum_\ell \int_{-\infty}^{\infty} [\partial_x \phi_\ell(x)]^2 dx$
where $v$ is the velocity of excitations, and $\phi_\ell(x)$ 
are \emph{chiral} Bose fields obeying the commutation relations
$[\phi_\ell(x),\phi_{\ell^\prime}(y)] =
i \pi \delta_{\ell \ell^\prime} \text{sgn}(x-y)$ \cite{bosonization}.
This representation is natural in the FQHE case \cite{chang03},
but also describes the quantum wires system, after unfolding
the decoupled (Bogolubov-transformed) right and left moving fields
\cite{gogolin94}.
Finally, the level and the leads are connected by tunneling terms
$H_{T} = \sum_\ell t_\ell d^{\dagger} \psi_\ell(0) + \text{H.c.}$
(characterized by amplitudes $t_\ell$),
and local interactions
$H_{U} = \sqrt{g} ( d^{\dagger}d - 1/2) \sum_\ell U_\ell
\partial_x \phi_\ell(0) / (2 \pi)$
(with strengths $U_\ell$).
The Fermi operators at the end of lead $\ell$
is related to the bosonic fields through
$\psi_{\ell}(0)=
\sqrt{D/2\pi v} \chi_\ell e^{i \phi_\ell (0)/\sqrt{g}}$,
where $\chi_\ell$ are Majorana fermions, and $D$ is the bandwidth.
In quantum wires $g$ is the LL interaction parameter
($g<1$ for repulsion, $g>1$ for attraction) \cite{bosonization};
for edges of FQHE at simple filling fraction $\nu$
we have $g=\nu$, assuming only electron tunneling,
i.e., a dot outside the quantum Hall bar, cf.\ Fig.~\ref{fig:geometry}(d)
\cite{chang03}.

In a recent work \cite{goldstein10b} we have demonstrated
that the side-coupled geometry of Fig.~\ref{fig:geometry}(b)
is related to the embedded configuration [Fig.~\ref{fig:geometry}(c)]
by a duality transformation $g \leftrightarrow 1/g$,
under which transmission maps onto reflection,
but the capacitance remains invariant.
We will thus continue to analyze the embedded system,
and later 
translate our results to the side-coupled geometry.

\section{Preliminary considerations} \label{sec:prelim}
Before going into an elaborate analysis 
we will try to use two limiting cases in order
to obtain some insights into the problem.
At $g=1$ (and $U_\ell=0$) we have just a noninteracting
resonant level, which has a constant low-temperature
capacitance, proportional to the inverse level width.
As tunneling into a LL is enhanced for $g>1$ \cite{bosonization,kane92},
this nonsingular behavior should apply there for any interaction strength.
Moreover, fermionic perturbative (in the electron-electron interaction)
renormalization group (RG) calculations \cite{nazarov03}
indicate that this simple behavior is not changed
qualitatively for $g$ smaller but close to unity.
Thus, interesting behavior is
expected only for sufficiently small $g$.

A deeper insight can be obtained from the 2CK problem,
describing a spin 1/2 impurity
coupled to two channels of spinful electrons \cite{hewson}.
Its anisotropic version is: 
\begin{multline}
\label{eqn:2ck}
 H_\text{2CK} = H_1 + H_2  + h_z S_z +
 S_z \left[ J_{1z} \sigma_{1z}(0) + J_{2z} \sigma_{2z}(0) \right] + \\
 S_+ \left[ J_{1xy} \sigma_{1-}(0) + J_{2xy} \sigma_{2-}(0) \right]/2 + \text{H.c.},
\end{multline}
where $h_z$ is a local magnetic field, the $J$'s are exchange couplings,
$S_\alpha$ are the components of the impurity spin ($\alpha=x,y,z$), and
$\sigma_{i\alpha}(x)$ are the components of spin density at $x$ of channel
$i=1,2$, governed by the noninteracting Hamiltonian $H_i$.
The latter can be bosonized, and expressed as
(using again a chiral representation)
$H_i = v/(4\pi) 
\int_{-\infty}^{\infty}
\left\{ \left[ \partial_x \phi_{i\rho}(x) \right]^2 +
\left[ \partial_x \phi_{i\sigma}(x) \right]^2 \right\} dx$,
featuring charge-spin ($\rho$-$\sigma$) separation.
Only the spin fields couple to the impurity, with
$\sigma_{i z}(0) = \partial_x \phi_{i \sigma} (0) / (\pi\sqrt{2})$ and
$\sigma_{i \pm} = [D/(2\pi v)] \exp[\mp i \sqrt{2} \phi_{i \sigma} (0)]$.
The 2CK
problem is equivalent to our original system,
provided that $g=1/2$,
with $\phi_{1\sigma}$ ($\phi_{2\sigma}$) being identified with
$\phi_L$ ($\phi_R$), and $S_{+}$ ($S_{-}$) being identified
with $d^\dagger$ ($d$), so $S_z \rightarrow d^\dagger d - 1/2$
\cite{fn:klein_factors}.
Under this mapping $J_{1z} = \sqrt{2g} U_L$, and
$J_{1xy} = 2 \sqrt{2\pi v/D} t_L$
(and similarly for the other channel/lead).
It is well known that the 2CK model exhibits a
logarithmically diverging impurity susceptibility
in the channel-symmetric case \cite{hewson},
which immediately implies a similar behavior for the capacitance
of our system at $g=1/2$.
Furthermore, for $g \ne 1/2$ our
system would be equivalent to a 2CK model with \emph{LL channels},
with spin LL
parameter $g_\sigma = 2g$ \cite{bosonization}.
Experimental realization of our system will thus enable the
investigation of this generalized 2CK problem.
For $g<1/2$ ($g_\sigma<1$) we might then suspect
that a behavior more singular than logarithmic may arise.
This expectation is indeed borne out by our subsequent calculations.

\section{General analysis} \label{sec:analysis}
We first apply the transformation
$\tilde{H} = \mathcal{U}^{\dagger} H \mathcal{U}$ with
$\mathcal{U}=e^{i \sqrt{g} \left( d^{\dagger}d - \frac{1}{2} \right)
[U_L \phi_L(0) + U_R \phi_R(0)] / (2 \pi v) }$, to
eliminate
the interaction term from the Hamiltonian,
at the cost of making the tunneling term
more complicated:
$ \tilde{H}_{T} = \xi^{-1} d^{\dagger} \sum_\ell 
        y_\ell \chi_\ell e^{i \sum_{\ell^\prime} 
	K_{\ell \ell^\prime} \phi_{\ell^\prime}(0)}
	+ \text{H.c.}$, 
where $y_\ell = t_\ell \sqrt{\xi/(2 \pi v)}$,
$K_{\ell \ell^\prime} =
[\delta_{\ell \ell^\prime}-gU_{\ell^\prime}/(2 \pi v)]/\sqrt{g}$,
and $\xi \sim D^{-1}$ is a short-time cutoff scale.
To proceed with RG analysis,
we need to include in $\tilde{H}$ another term
(invariant under $\mathcal{U}$),
$H_{LR} = 
 2 \xi^{-1} y_{LR} \left( d^\dagger d - \textstyle\frac{1}{2} \right)
 \chi_L \chi_R e^{i[\phi_L(0)-\phi_R(0)]/\sqrt{g}} + \text{H.c.}$, 
describing inter-lead cotunneling with dimensionless
amplitude $y_{LR}$,
generated by RG via virtual processes.

Using the scaling dimensions and operator product expansions of the
various terms in the Hamiltonian, we can derive RG
equations for the various terms
(Here $\delta_\ell \equiv g U_\ell/(2 \pi v)$,
$\delta_\pm \equiv (\delta_L \pm \delta_R)/2$, and $\tilde{\ell}=R$
for $\ell=L$ and vice-versa) \cite{bosonization,cardy}:
\begin{align}
 \label{eqn:rg_yell}
 \frac {dy_\ell} { d \ln \xi} & =
 \left[ 1 - \frac{(1 - \delta_\ell)^2}{2g}
 - \frac{(\delta_{\tilde{\ell}} )^2}{2g} \right] y_\ell
 + y_{\tilde{\ell}} y_{LR}
 , \\
  \label{eqn:rg_ylr}
 \frac {dy_{LR}} { d \ln \xi} & =
 \left( 1 - \frac{1}{g} \right) y_{LR}
 + y_{L} y_{R}
 , \\
 \label{eqn:rg_dp}
 \frac {d\delta_+} { d \ln \xi} & =
 \left( 1 - 2\delta_+ \right)
 \left(y_L^2 + y_R^2 \right)
 , \\
 \label{eqn:rg_dm}
 \frac {d\delta_-} { d \ln \xi} & =
 \left( y_L^2-y_R^2 \right) - 2 \left(y_L^2 + y_R^2 \right) \delta_-
 .
\end{align}
These equations show that
left-right asymmetry is relevant in general.
Thus, in $L$-$R$ asymmetric case the system will end up
(below an energy scale $T_{a}$,
behaving as $\sim (t_L-t_R)^{2g /
[ 2g - (1 - \delta_{+})^2 - \delta_{+}^2 ] }$
for $U_L=U_R$)
in a fixed point where the dot is strongly-coupled to one of the leads.
The capacitance 
is thus the same as that of a dot coupled to one lead
[Fig.~\ref{fig:geometry}(a)], which does
not exhibit any LL-type power-law behavior, as discussed above.
As for transport, we effectively have
a simple tunnel-junction between two leads, corresponding
to the dot and the weakly-coupled lead in the original system
\cite{kane92}. 

We will therefore consider the more interesting symmetric case,
$t_L=t_R=t$ and $\delta_L=\delta_R=\delta_{+}$.
It should be noted that this situation arises naturally in
the side-coupled geometry [Fig.~\ref{fig:geometry}(b)],
where the two channels are the left- and right-movers
in the same lead \cite{goldstein10b}.
In the embedded [Fig.~\ref{fig:geometry}(c)] or FQHE configurations,
$L$-$R$ asymmetry can be controlled via voltages applied to
the electrodes controlling the tunnel junctions.
Since the channel symmetric case features the highest on-resonance
transmission \cite{kane92} 
(see also below),
changing tunnel-junction parameters while monitoring conductance
through the system can help tune the system into $L$-$R$ symmetry.

Although we are primarily interested in the capacitance
here, we will start with a discussion of the $L$-$R$
conductance \cite{kane92}, 
in order to make the presentation more self-contained.
Off-resonance ($\varepsilon_0 \ne 0$), the flow of $y_\ell$ is
stopped when $\xi^{-1} \sim |\varepsilon_0|$. From this point on,
the level is locked into one of its two possible states
[nearly occupied (empty), for negative (positive) $\varepsilon_0$].
The system thus behaves as a tunnel junction,
whose strength $y_{LR}/\xi \sim t^2/\varepsilon_0$ for
large enough $\varepsilon_0$.
This cotunneling process will be relevant for $g>1$
and irrelevant for $g<1$ \cite{kane92}.
On resonance $t$
is the leading low energy process.
It is relevant (when small enough) for
$g > g_c = \left[(1-\delta_{+})^2+\delta_{+}^2\right]/2$,
irrelevant otherwise.
Whenever any of these two terms is relevant, tunneling through 
the level enables perfect conductance for small temperature
and bias voltage.

For strong level-lead coupling,
the dot-lead interaction rapidly converges to its fixed point value
$\delta_{+} = 1/2$
(cf.\ Eq.~(\ref{eqn:rg_dp}); At $g=1/2$, this is the 
Emery-Kivelson point of the equivalent 2CK problem \cite{komnik03}).
Working in terms of the symmetric and anti-symmetric combinations of
the bosonic fields
$\phi_{s/a}(x)=(\phi_L(x) \pm \phi_R(x))/\sqrt{2}$,
the former then does not couple to the dot. 
On resonance, $\tilde{H}_{T}$
is again dominant.
Using the equivalent 2CK spin operators
\cite{fn:klein_factors},
it acquires the form
$\sim S_x \cos[\phi_a(0)/\sqrt{2g}]$
so that $S_x$ assumes one of its possible
values ($\pm 1/2$), and $\phi_a(0)$ is confined to one of
the resulting minima of this term.
One can then expand the partition function
in the amplitude for tunneling events of $\phi_a(0)$
between these minima (instantons),
and find that it is relevant for $g<1/4$ and irrelevant for $g>1/4$.
Off resonance only the cotunneling process is important at low energies,
as discussed above. In the spin notation it becomes
$\sim S_z \cos[\phi_a(0)\sqrt{2/g}]$, so now instantons are relevant
for $g<1$ and irrelevant for $g>1$ \cite{kane92}.
Both on and off resonance,
relevant instantons imply suppressed transmission and vice-versa.

\begin{table}
\caption{ \label{tbl:phases}
Definition of the different phases of the system,
corresponding to the phase diagram, Fig.~\ref{fig:phase_diagram}.
$G_\text{on(off)}$ is the zero temperature
on-resonance (off-resonance) linear conductance.
$(T/T_K)^{2g-1}/T_K$ becomes $\ln(T/T_K)/T_K$ at $g=1/2$.
In the side-coupled system $G$
is replaced by $e^2/h - G$,
and $g$ by $1/g$ \cite{goldstein10b}.
See the text for further details.
}
\begin{ruledtabular}
\begin{tabular}{cccc}
 Phase  & $G_\text{on}$ & $G_\text{off}$ & Capacitance               \\ \hline
 SC-OFF & $e^2/h$       & $e^2/h$        & $\sim 1/T_K$              \\
 SC-ON1 & $e^2/h$       & $0$            & $\sim 1/T_K$              \\
 SC-ON2 & $e^2/h$       & $0$            & $\sim (T/T_K)^{2g-1}/T_K$ \\
 WC     & $0$           & $0$            & $\sim 1/T$    
\end{tabular}
\end{ruledtabular}
\end{table}

\begin{figure}
\includegraphics[width=8cm,height=!]{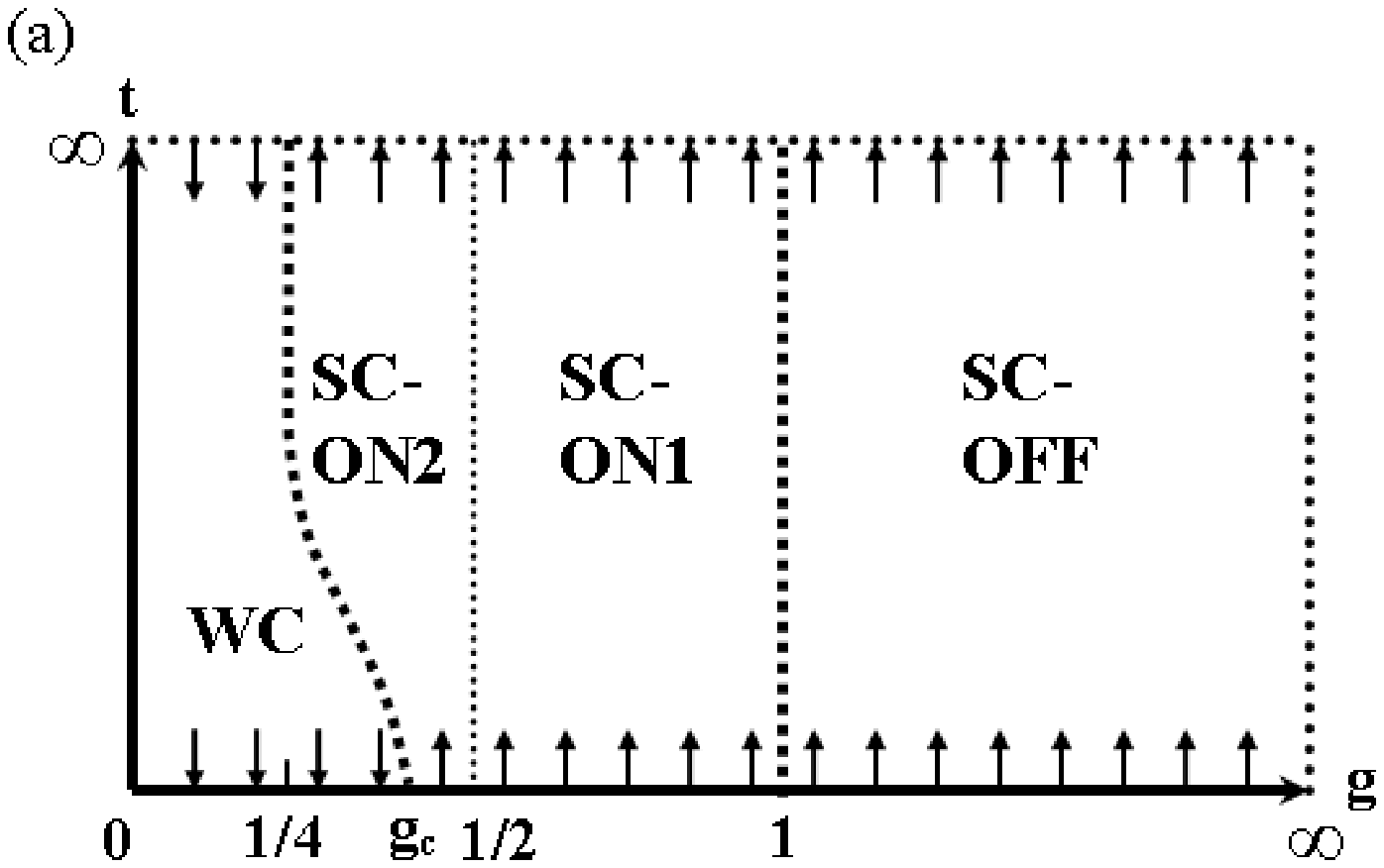}
\includegraphics[width=8cm,height=!]{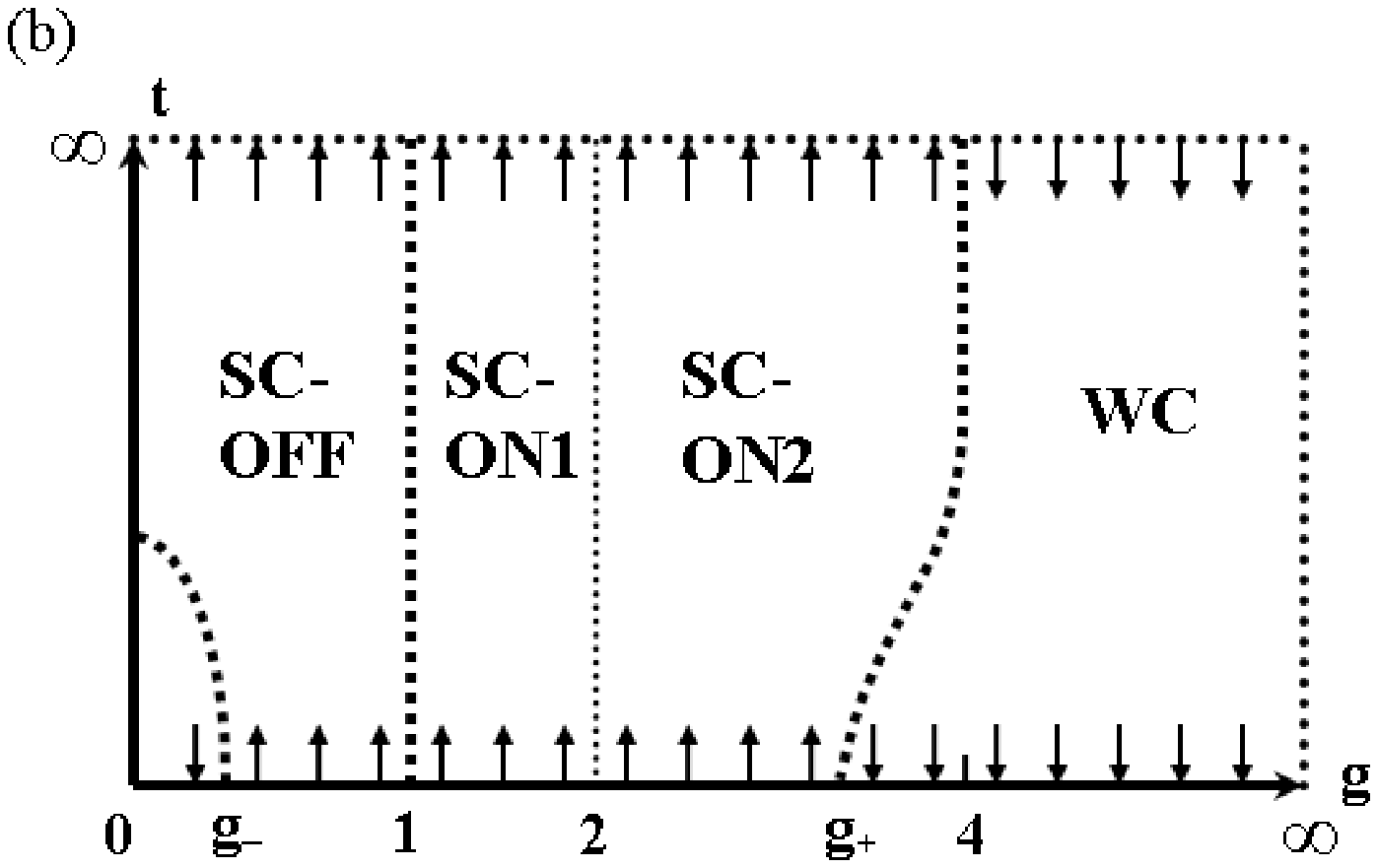}
\caption{\label{fig:phase_diagram}
Zero temperature phase diagram (phases defined in Table~\ref{tbl:phases})
and on-resonance RG flow,
projected on the $g$-$t$ plane:
(a) embedded geometry; (b) side-coupled geometry.
$g_{\pm}$ are the two roots of $g+(1-\delta_{+})^2/g=4$ \cite{goldstein10b,fn:g_lt_gm}.
See the text for further details.
}
\end{figure}

To summarize, we have three phases regarding low-temperature $L$-$R$
conductance:
(i) The weak-coupling phase (WC), with suppressed low-energy conductance
both on and off resonance;
(ii) The strong-coupling on-resonance phase (SC-ON), with good on-resonance
but suppressed off-resonance transmission;
(iii) The strong-coupling off-resonance phase (SC-OFF), with good on-
\emph{and} off-resonance transport.
These phases, together with the corresponding picture
for the side-coupled geometry, are summarized in Table~\ref{tbl:phases} and
Fig.~\ref{fig:phase_diagram} \cite{goldstein10b,fn:g_lt_gm}.
Phase SC-ON is further divided into two regions according to the
behavior of the capacitance, as we now discuss.

For the level population, only the $t$ process is important
(as cotunneling does not change the dot occupancy). Interesting
behavior can thus arise only for small $|\varepsilon_0|$.
In the WC phase the level is effectively decoupled from the reservoirs
at low energies, thus featuring discontinuous dependence
of the population on $\varepsilon_0$ (as the latter crosses the Fermi
energy), or equivalently, $1/T$ divergence of the capacitance.
In the SC phases continuous behavior is expected.
Let us now analyze this behavior in more details,
using methods similar to those employed in
Refs.~\onlinecite{gogolin94}.

By the Kubo formula, the (dynamic) level capacitance
is determined by the retarded population-population correlation function.
The latter can be obtained via analytical continuation from the imaginary
time Green function, which, in the spin notation, is 
$\chi_{zz}(\tau) \equiv - \langle \hat{T}_\tau S_z(\tau) S_z(0) \rangle=
- \text{Tr} [ e^{(|\tau|-1/T)\tilde{H}} S_z
e^{-|\tau|\tilde{H}} S_z ] /Z$,
with $\hat{T}_\tau$ the time-ordering operator and
$Z$ the partition function
(note that
$\tilde{S}_z = \mathcal{U}^\dagger S_z \mathcal{U} = S_z$).
Near the strong coupling fixed point
$\tilde{H}_{T} \sim S_x \cos[\phi_a(0)/\sqrt{2g}]$, as discussed above.
Hence, $S_z \tilde{H}_{T} S_z = - \tilde{H}_{T}/4$.
Since the different terms in $\tilde{H}_{T}$
change the population of one of the leads by $\pm 1$, we also have
$\mathcal{V}^\dagger \tilde{H}_T \mathcal{V} = - \tilde{H}_{T}$,
where
$\mathcal{V}=e^{i \pi (N_L - N_R) }$,
with $N_\ell$ the electron number operator in lead $\ell$.
The other terms in the Hamiltonian commute with both $S_z$
and  $\mathcal{V}$.
Thus, the charge susceptibility can be written as
$\chi_{zz}(\tau) =
-\langle \hat{T}_\tau \mathcal{V}^\dagger(\tau) \mathcal{V}(0) \rangle/4$.
Since $N=N_L+N_R$ is constant, we get
$\ln \chi_{zz}(\tau) \sim
- 2\pi^2 \langle \hat{T}_\tau N_L(\tau) N_L(0) \rangle$.
At the strong-coupling fixed point the two leads are well-connected,
so the latter correlator 
is simply the correlation function
of the occupation of one half of an infinite wire,
leading to
$\chi_{zz}(\tau) \sim \tau^{-2g}$. Fourier-transforming, we
find that $\chi_{zz}(\omega)$ is regular for small frequencies
for $g>1/2$, i.e., it goes to a constant for vanishing $\omega$.
This constant will be proportional to $1/T_K$, with
the strong-coupling scale $T_K$ (``Kondo temperature'')
behaving as $\sim t^{2g /
[ 2g - (1 - \delta_{+})^2 - \delta_{+}^2 ] }$
for small $t$ [In the noninteracting case ($g=1$ and $U_\ell=0$)
$T_K \sim t^2$ is simply the level width].
However, for $g<1/2$, the dynamic capacitance
will have a power-law singularity $\chi(\omega) \sim \omega^{2g-1}$.
The power-law 
will be smeared by finite level energy,
temperature, lead-length $L$, or channel asymmetry.
The static capacitance will thus vary as $\sim \Lambda^{2g-1}$
with $\Lambda$ the largest of these scales
(i.e., $|\varepsilon_0|$, $T$, $v/L$, or $T_{a}$, respectively)
\cite{fn:perturb}.
At $g=1/2$, the power-law will become a logarithm,
as expected from the 2CK analogy discussed above.
The different behaviors are again summarized in Table~\ref{tbl:phases} and
Fig.~\ref{fig:phase_diagram}.

\section{Conclusions} \label{sec:conclude}
To conclude, we have shown how the differential capacitance of
a level coupled to the ends of two LL wires, or side-coupled to
a single wire, can exhibit power-law divergence at low temperatures,
which enables a measurement of the LL parameter $g$.
This might help to resolve the puzzle regarding tunneling
measurements in FQHE edges at $\nu=1/3$ filling,
which show power-law temperature dependence
deviating from the predicted one \cite{chang03}. 
It should be noted that in a related large-dot system power-law
behavior of the capacitance was predicted to occur only in a subleading
term, and would thus be harder to detect
\cite{kolomeisky02,kakashvili03}.
In addition, the behavior found here reveals
generalized 2CK physics. Hence, it creates 
an alternative route for its realization in the laboratory,
over existing possibilities \cite{matveev95,potok07}.

\begin{acknowledgments}
We would like to thank Y. Gefen, I.V. Lerner, A. Schiller, and I.V. Yurkevich
for useful discussions.
Support by the Adams Foundation of the Israel Academy
of Science and by the Israel Science Foundation (Grant 569/07) is
gratefully acknowledged.
\end{acknowledgments}


\begin{thebibliography}{99}

\bibitem{hewson}
D.L. Cox and A. Zawadowski,
Adv. Phys. \textbf{47}, 599 (1998).

\bibitem{bosonization}
T. Giamarchi, \textit{Quantum Physics in One Dimension}
(Oxford University Press, Oxford, 2003).

\bibitem{chang03}
X.-G. Wen, \textit{Quantum Field Theory of Many-Body Systems}
(Oxford University Press, Oxford, 2004);
A.M. Chang, \rmp \textbf{75}, 1449 (2003).

\bibitem{kane92} C.L. Kane and M.P.A. Fisher, \prl \textbf{68}, 1220 (1992);
\prb \textbf{46}, R7268 (1992); \textbf{46}, 15233 (1992).

\bibitem{nazarov03}
Yu.V. Nazarov and L.I. Glazman, \prl \textbf{91}, 126804 (2003);
D.G. Polyakov and I.V. Gornyi, \prb \textbf{68}, 035421 (2003);
I.V. Lerner, V.I. Yudson, and I.V. Yurkevich, \prl
\textbf{100}, 256805 (2008).


\bibitem{komnik03}
A. Komnik and A.O. Gogolin, \prl \textbf{90}, 246403 (2003);
\prb \textbf{68}, 235323 (2003).

\bibitem{goldstein10b}
M. Goldstein and R. Berkovits,
\prl \textbf{104}, 106403 (2010).

\bibitem{furusaki02}
A. Furusaki and K.A. Matveev, \prl \textbf{88}, 226404 (2002).

\bibitem{kolomeisky02}
E.B. Kolomeisky, R.M. Konik, and X. Qi, \prb \textbf{66}, 075318 (2002);
E.B. Kolomeisky, M. Timmins, and R.M. Kalas, 
\textit{ibid.} \textbf{74}, 245124 (2006).

\bibitem{kakashvili03}
P. Kakashvili and H. Johannesson,
\prl \textbf{91}, 186403 (2003);
Europhys. Lett. \textbf{79}, 47004 (2007).

\bibitem{sade05}
M. Sade
, Y. Weiss, M. Goldstein, and R. Berkovits,
\prb \textbf{71}, 153301 (2005);
Y. Weiss, M. Goldstein, and R. Berkovits, 
\textit{ibid.}
\textbf{75}, 064209 (2007); \textbf{76}, 024204 (2007);
\textbf{77}, 205128 (2008).

\bibitem{lehur05}
K. Le Hur and M.-R. Li, \prb \textbf{72}, 073305 (2005).

\bibitem{wachter07}
P. W\"{a}chter, V. Meden, and K. Sch\"{o}nhammer,
\prb \textbf{76}, 125316 (2007).

\bibitem{bishara08}
G.A. Fiete, W. Bishara, and C. Nayak,
\prl \textbf{101}, 176801 (2008);
\prb \textbf{82}, 035301 (2010).

\bibitem{goldstein09}
M. Goldstein, Y. Weiss, and R. Berkovits,
Europhys. Lett. \textbf{86}, 67012 (2009);
Physica E \textbf{42}, 610 (2010).

\bibitem{buttiker93}
M. B\"{u}ttiker,  A. Pr\^{e}tre, and H. Thomas,
\prl \textbf{70}, 4114 (1993);
S.E. Nigg, R. L\'{o}pez, and M. B\"{u}ttiker, 
\textit{ibid.} \textbf{97}, 206804 (2006).

\bibitem{berman99}
D. Berman, N.B. Zhitenev, R.C. Ashoori, and M. Shayegan,
\prl \textbf{82}, 161 (1999).

\bibitem{gabelli06}
J. Gabelli \textit{et al.},
Science \textbf{313}, 499 (2006).

\bibitem{hamamoto10}
Y. Hamamoto, T. Jonckheere, T. Kato, and T. Martin,
\prb \textbf{81}, 153305 (2010).

\bibitem{mora09}
C. Mora and K. Le Hur, Nat. Phys. \textbf{6}, 697 (2010).

\bibitem{gogolin94}
A.O. Gogolin and N.V. Prokof'ev, \prb \textbf{50}, 4921 (1994);
M. Fabrizio and A.O. Gogolin, 
\textit{ibid.} \textbf{50}, 17732 (1994);
\textbf{51}, 17827 (1995).

\bibitem{fn:klein_factors}
The Majorana fermions can be absorbed into these definitions to make
the spin variables commute with the lead degrees of freedom \cite{komnik03}.


\bibitem{cardy}
J. Cardy, \textit{Scaling and Renormalization in Statistical Physics}
(Cambridge University Press, Cambridge, 1996).

\bibitem{fn:g_lt_gm}
In the lower-left corner of Fig.~\ref{fig:phase_diagram}(b)
$L$-$R$ backscattering might become
dominant over level-wire tunneling even on resonance.
This should result in suppressed transmission together with $\sim 1/T$ capacitance.

\bibitem{fn:perturb}
Perturbative calculations \cite{wachter07} also predict power-law behavior,
but with a different power and in a different region of parameter space.

\bibitem{matveev95}
K.A. Matveev, \prb \textbf{51}, 1743 (1995).

\bibitem{potok07}
R.M. Potok \textit{et al.},
Nature \textbf{446}, 167 (2007).

\end{thebibliography}
\end{document}